# Discovery of segmented Fermi surface induced by Cooper pair momentum

Zhen Zhu[1+], Michał Papaj[2+], Xiao-Ang Nie[1], Hao-Ke Xu[1], Yi-Sheng Gu[1], Xu Yang[1], Dandan Guan[1], Shiyong Wang[1], Yaoyi Li[1], Canhua Liu[1], Jianlin Luo[3], Zhu-An Xu[4], Hao Zheng[1*], Liang Fu[2*], Jin-Feng Jia[1*]

1. School of Physics and Astronomy, Key Laboratory of Artificial Structures and Quantum Control (Ministry of Education), Shenyang National Laboratory for Materials Science, Tsung-Dao Lee Institute, Shanghai Jiao Tong University, Shanghai 200240, China
2. Department of Physics, Massachusetts Institute of Technology, Cambridge, MA 02139, USA
3. Beijing National Laboratory for Condensed Matter Physics, Institute of Physics, Chinese Academy of Sciences, Beijing 100190, China
4. Department of Physics, Zhejiang University, Hangzhou 310027, Zhejiang, China

Correspondence to: haozheng1@sjtu.edu.cn; liangfu@mit.edu; jfjia@sjtu.edu.cn

**Since the early days of Bardeen-Cooper-Schrieffer theory, it has been predicted that a sufficiently large supercurrent can close the energy gap in a superconductor and creates gapless Bogoliubov quasiparticles through the Doppler shift of quasiparticle energy due to the Cooper pair momentum(*1*) (Fig. 1A). In this gapless superconducting state, zero-energy quasiparticles reside on a segment of the normal state Fermi surface, while its remaining part is still gapped (see Fig. 1B). The finite density of states of field-induced quasiparticles, known as the Volovik effect(*2*), has been observed in tunneling(*3, 4*) and specific heat measurements(*5, 6*) on d- and s-wave superconductors. However, the segmented Fermi surface of a finite-momentum state carrying a supercurrent has never been detected directly. Here we use quasiparticle interference (QPI) technique to image field-controlled Fermi surface of $Bi_2Te_3$ thin films proximitized by the superconductor $NbSe_2$. By applying a small in-plane magnetic field, a screening supercurrent is induced which leads to finite-momentum pairing on topological surface states of $Bi_2Te_3$. Our measurements and analysis reveal the strong impact of finite Cooper pair momentum on the quasiparticle spectrum, and thus pave the way for STM study of pair density wave and FFLO states in unconventional superconductors(*7, 8*).**

Our platform of choice, shown in Fig.1C, is thin films of $Bi_2Te_3$, a quintessential topological insulator (TI)(*9, 10*), grown by molecular beam epitaxy on top of a bulk crystal $NbSe_2$, an s-wave superconductor. Proximitized TIs have previously been investigated for the presence of finite momentum pairing states(*11*). The superconducting proximity effect from $NbSe_2$ induces a hard gap in $Bi_2Te_3$ at zero field, as demonstrated in previous STM measurements(*12*). This material combination provides a synergy ideal for creating and detecting the gapless superconducting state. $NbSe_2$ is a clean s-wave superconductor with a large superconducting gap and long penetration depth $\lambda$ over 100 nm(*13*). Importantly, this means that a small in-plane magnetic field that doesn't induce appearance of vortices in the measurement region can cause a spatially varying complex order parameter $\Delta(\boldsymbol{r}) = \Delta_0 \exp(i \boldsymbol{q} \cdot \boldsymbol{r})$ on the surface. Here, the Cooper pair momentum $q \sim B_{ext}\lambda$ is in the direction perpendicular to the field $\boldsymbol{q} \perp \boldsymbol{B}_{ext}$. To probe this finite momentum $q$ we use the $Bi_2Te_3$ film under proximity effect, which has a simple Fermi pocket surrounding Γ point(*14*) that originates from the topological surface states. The proximity induced gap is reduced and thus only a small magnetic field is needed to significantly affect the quasiparticle spectrum without strongly impacting the parent superconductor. Moreover, the spin-helical nature of these states introduces a new twist. Recent theory predicts that when gapless quasiparticles in a proximitized topological insulator are confined into a quasi-1D channel, topological superconductivity hosting Majorana end states may be formed(*15*).

In the insets of Fig.1C we present the topography of our thin film with regions of varying thickness, and the atomic resolution image of $Bi_2Te_3$ lattice, demonstrating its high quality. The properties of $Bi_2Te_3$ films are significantly affected by the number of the quintuple layers (QL), 1 nm thick basic building blocks of this crystal. Topological surface states are formed at the thickness above 3 QL (*16*), while our previous results show that proximity-induced superconducting energy gap is present in the top surface for films up to 11 QL thick (*12*). To optimize the superconductivity in the topological surface states, we perform all our measurements on 4 QL area of the sample, highlighted by the white dashed square of area 120x120 $nm^2$ away from the step edges. Fig.1D and 1E show a set of differential conductance dI/dV curves along the line cut inside this region. They display a high degree of spatial uniformity over a wide range of energy scales from 0.1eV to -0.43eV (Fig.1D). We identify conduction band minimum and valence band maximum from dI/dV peaks at -0.3eV and -0.07 eV respectively, establishing the Fermi level position at about 360 meV above the surface Dirac point. Near the Fermi level, we observe a hard, U-shaped superconducting energy gap $\Delta \approx 0.5$ meV at zero magnetic field, with no visible in-gap features across the line cut.

We now apply in-plane vector magnetic field to the thin film and measure the differential conductance to investigate the gapless superconducting state. Considering the strong hexagonal warping effect of $Bi_2Te_3$ surface states(*17*), we orient the magnetic field along two different high symmetry directions ΓK and ΓM. The dI/dV spectra, presented in Fig.2, reveal a remarkably rich set of in-gap features. As the magnetic field is increased in small steps of 10mT, multiple distinct peaks and shoulders appear and change rapidly. The in-gap spectrum also depends on the field direction. This contrasts with the surface of pristine $NbSe_2$ (see SM), where the in-gap spectrum is featureless under the same magnetic field, displaying a hard gap with only minimal changes to the coherence peaks. Moreover, these observations contrast sharply with the tunneling spectra of conventional superconductors such as aluminum or lead, where magnetic field also causes the filling of the superconducting gap in a featureless manner(*4*) as magnetic impurities do(*18*). To understand the microscopic origin of the observed tunneling spectra, we perform theoretical calculation of the density of states at various field-induced Cooper pair momenta. Remarkably, all distinctive in-gap features and their evolution with the field in both directions are reproduced by our calculation based on the established model Hamiltonian for $Bi_2Te_3$ surface states including hexagonal warping. We also highlight that if the effects mentioned were to origin from the Zeeman effect instead of the screening supercurrent, the Zeeman energy to close the superconducting gap at 20 mT would have to be 25 meV/T, translating to the effective g-factor of few hundreds. This is an unrealistic, order of magnitude increase in comparison to the values observed in various topological materials(*19–22*).

Our calculations thus show the emergence of a segmented Fermi surface(*15, 23*) in the superconducting state above a critical field. The evolution of spectral function at the Fermi level is depicted in the animation in SM. At zero field, the normal state Fermi surface, a six-pointed star, is fully gapped by s-wave pairing. When the magnetic field is applied, the screening supercurrent induces a finite Cooper pair momentum and causes depairing of quasiparticles moving parallel to the supercurrent. As a result, the quasiparticle energy dispersion becomes tilted. As the depairing energy increases with the field and becomes comparable to the superconducting gap, Bogoliubov quasiparticles appear at zero-energy and their loci in momentum space defines a Fermi surface. Importantly, this quasiparticle Fermi surface is formed by electron and hole segments of the normal state Fermi surface around the pair-breaking region, as evidenced by the spectral function.

To detect the segmented Fermi surface directly, we scan the constant energy local density of states over the whole region of interest and perform a Fourier transform to obtain the quasiparticle interference patterns (QPI)(24, 25). At high energies far above the superconducting gap, we observe strong and equal intensities at six segments, which are symmetrically placed along three equivalent ΓM directions. This QPI pattern is independent of the direction or magnitude of magnetic field, and similar to those observed in $Bi_2Te_3$ without superconductivity(26–28).

However, the QPI pattern becomes drastically different at the energies inside of the superconducting gap. Pairs of real and momentum space images at zero energy are presented in Fig. 3 for six different orientations of magnetic field along high symmetry directions at $B = 40$ mT. In the real space images we observe one-dimensional standing wave patterns, whose orientation changes with the magnetic field direction. These real-space patterns yield two distinct classes of Fourier images, showing respectively 2 and 4 high-intensity segments out of the original 6. When the field is applied along ΓK, two bright segments are found along the perpendicular ΓM direction, while the remaining four are strongly suppressed. When the field is directed along ΓM, the corresponding two segments are dark, while the other four segments are bright.

These QPI patterns can be understood directly from the picture of segmented Fermi surface in the superconducting state under an in-plane magnetic field. Due to the direction-dependent depairing effect, only a portion of the normal state Fermi surface becomes gapless, and thus only a subset of hotspots in this gapless segment is activated for scattering at zero energy. To illustrate this, in Fig. 4G-I we present the spectral function of the normal state, superconducting state with magnetic field along ΓK, and superconducting state with magnetic field along ΓM, respectively. In normal state, we see a hexagonally warped contour, which gives rise to six symmetric segments in the QPI pattern (Fig. 4A) associated with scattering between the hotspots at the neighboring tips of the star. In the superconducting state at zero field, there is no Fermi surface due to the hard gap. Nevertheless, it re-emerges at B=40 mT due to the filling of the gap with quasiparticle states. However, this new Fermi surface only consists of segments of the normal state Fermi surface, whose size and location are controlled by the field strength and orientation (Fig. 4 H, I). Then, scattering between the available hotspots on the segmented Fermi surface gives rise to those bright segments in the observed QPI pattern (Fig. 4B, C), as indicated by the marked momentum transfer vectors $Q_i$. In contrast, those gapped hotspots cannot participate in quasiparticle scattering processes, leading to a suppression of QPI intensity at corresponding wavevectors.

Interestingly, scattering between Bogoliubov quasiparticles on the segmented Fermi surface is strongly dependent on the superconducting coherence factors. Zero-energy quasiparticles having momentum $k \cdot B > 0$ and $k \cdot B < 0$ are symmetric and antisymmetric superposition of electron and hole states, respectively. In the presence of nonmagnetic impurity, scattering between the two opposite branches is constructively enhanced, while scattering within each branch is destructively suppressed. For this reason, even though the wavevector $Q_1$ connects the same pair of hotspots in both Fig.4H and Fig.4I, quasiparticle scattering at this wavevector is present in the former case but suppressed by the coherence factor in the latter. This can be clearly seen in Fig.4B and 4C, thus proving the identity of gapless excitations as Bogoliubov quasiparticle rather than normal electron. To further substantiate the above theoretical analysis, we perform a full numerical simulation of proximitized topological surface state under in-plane magnetic field. By using recursive Green functions (details in SI) we calculate the local density of states in the presence of random disorder and construct its Fourier image. Our numerical QPI patterns, shown in Fig.4 (D-F), show a remarkably good agreement with the experimental data.

To summarize, our energy- and momentum-resolved QPI maps provide the first direct evidence of the quasiparticle Fermi surface in finite-momentum gapless superconducting state of $Bi_2Te_3/NbSe_2$. The gap closing mechanism based on the screening supercurrent also facilitates a new way of controlling topological phase of the superconductor, possibly leading to new platforms for studying Majorana bound states.

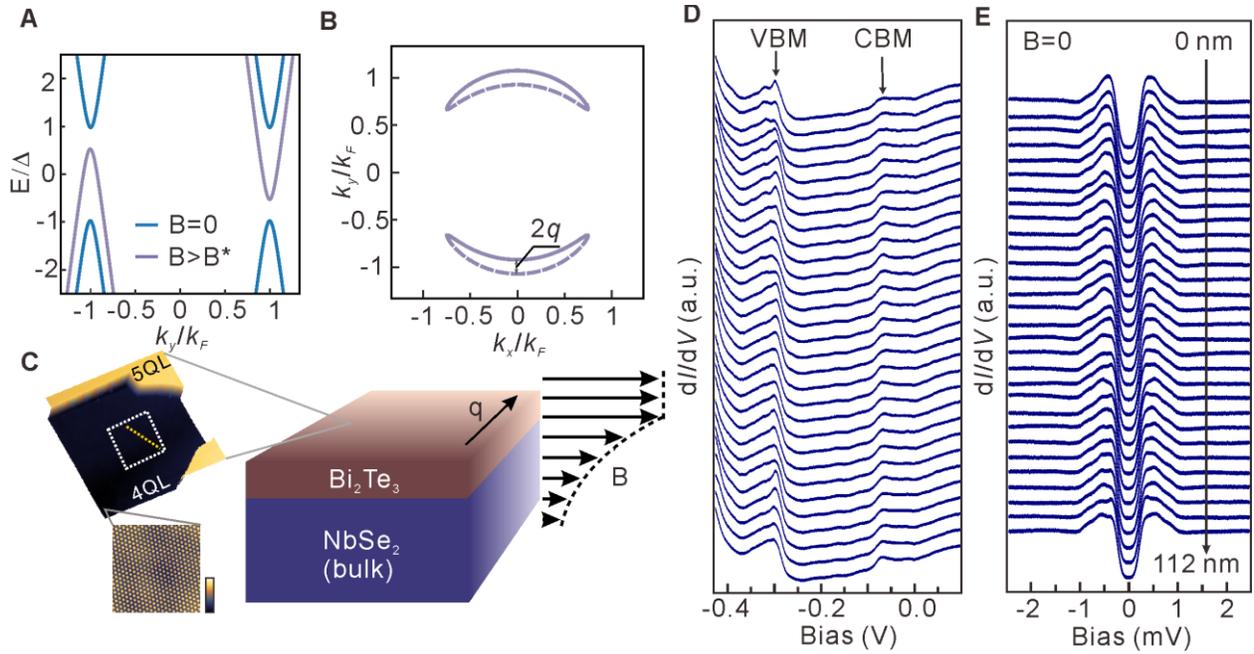

**Fig. 1 Topography and characterization of proximitized Bi$_2$Te$_3$ thin films.** (A) Quasiparticle dispersion around superconducting gap without magnetic field (blue curve) and with magnetic field above B* (purple curve) that induces gapless superconducting state. (B) Schematic depiction of a segmented Fermi surface. (C) Structure of the samples with few quintuple layers of Bi$_2$Te$_3$ on top of NbSe$_2$. Insets show the topography of a large area of the thin film with regions of varying thickness and atomic resolution image of Bi$_2$Te$_3$ lattice. (D) Wide energy range differential conductance dI/dV spectra taken across yellow line cut in (C). The set point is 0.1V, 0.5nA. CBM (VBM) stands for the conduction band minimum (valence band maximum) of the Bi$_2$Te$_3$ bulk bands. Fermi energy is about 70 meV above CBM, which implies the surface Dirac point is about 360 meV below Fermi level. (E) The proximity induced superconducting gap along the same line cut as in (C). The set point is 2.5mV, 0.5nA. From the position of the coherence peaks we determine superconducting gap $\Delta \approx 0.5$ meV.

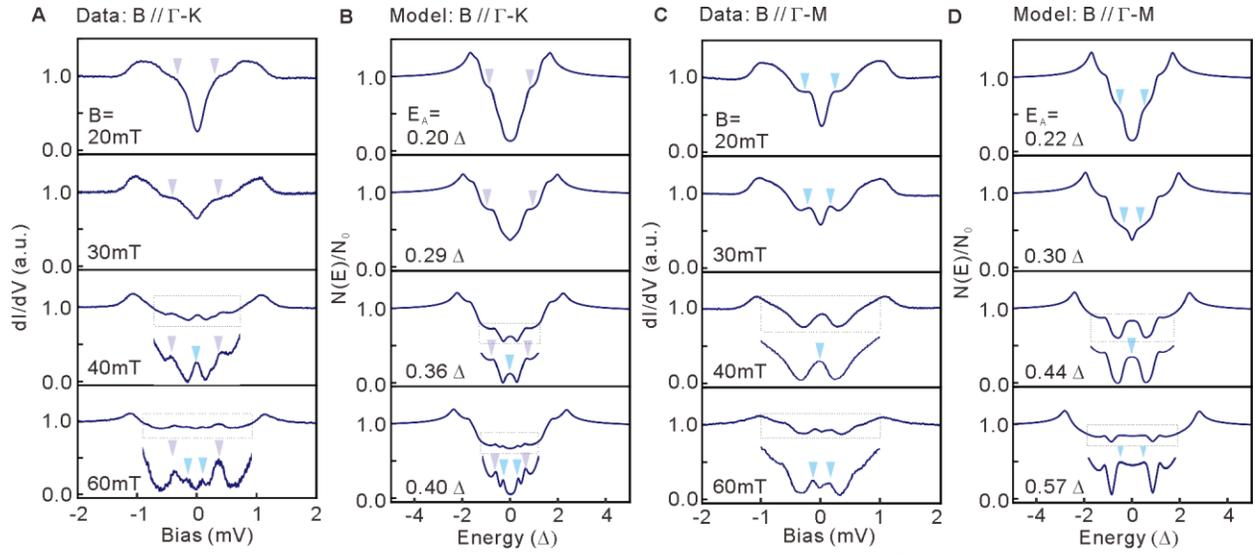

**Fig. 2 Density of states under in-plane magnetic field.** (A) Differential conductance dI/dV spectra and (B) theoretical density of states curves for increasing magnetic field along ΓK direction. (C) Differential conductance dI/dV spectra and (D) theoretical density of states curves for increasing magnetic field along ΓK direction. For both (A) and (C) the spectra set points are 2mV and 0.5nA. The characteristic energy scale for theoretical calculations is $E_A = evA_{x,y}$.

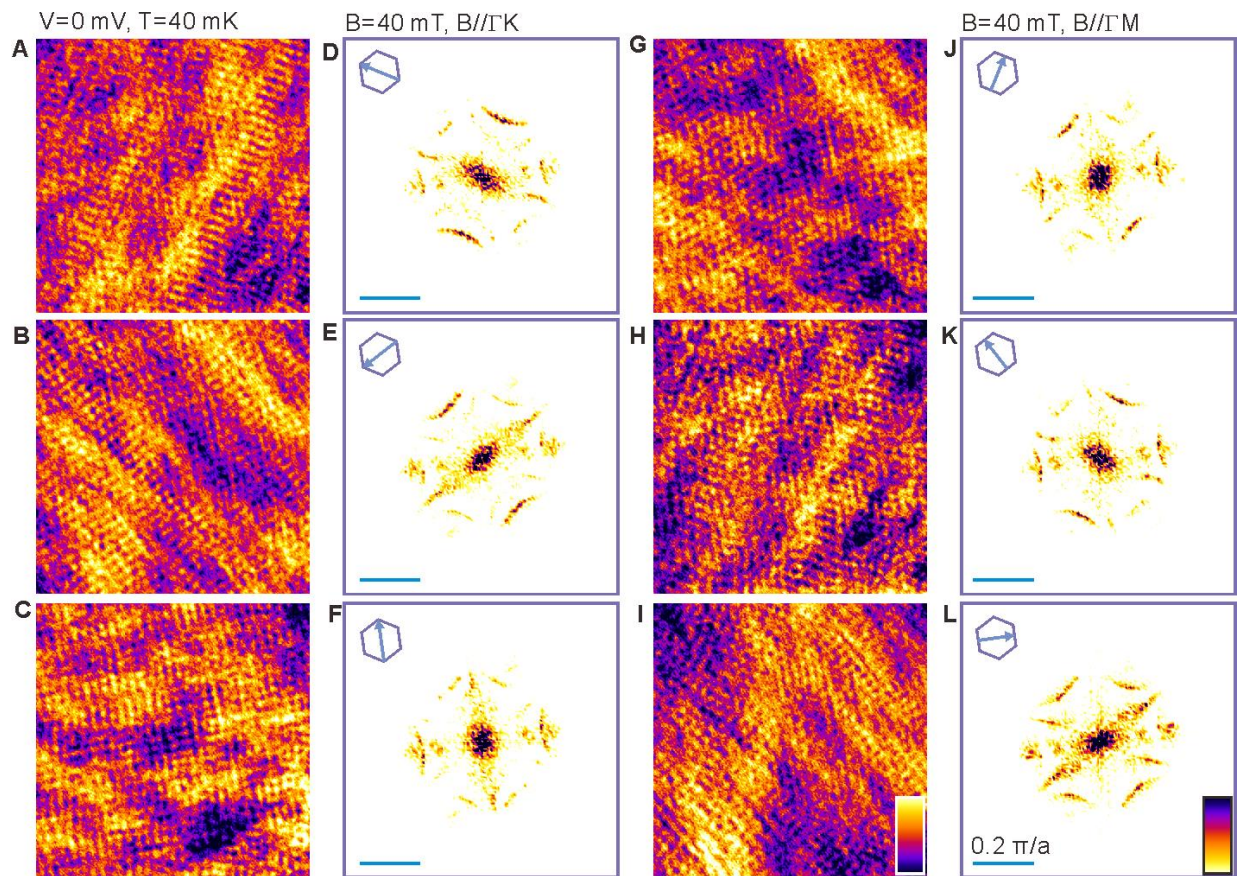

**Fig. 3 Real and momentum space quasiparticle interference patterns for 6 different orientations of the in-plane magnetic field.** The momentum space patterns display either 2 or 4 bright segments for magnetic field along Γ$K$ or Γ$M$, respectively.

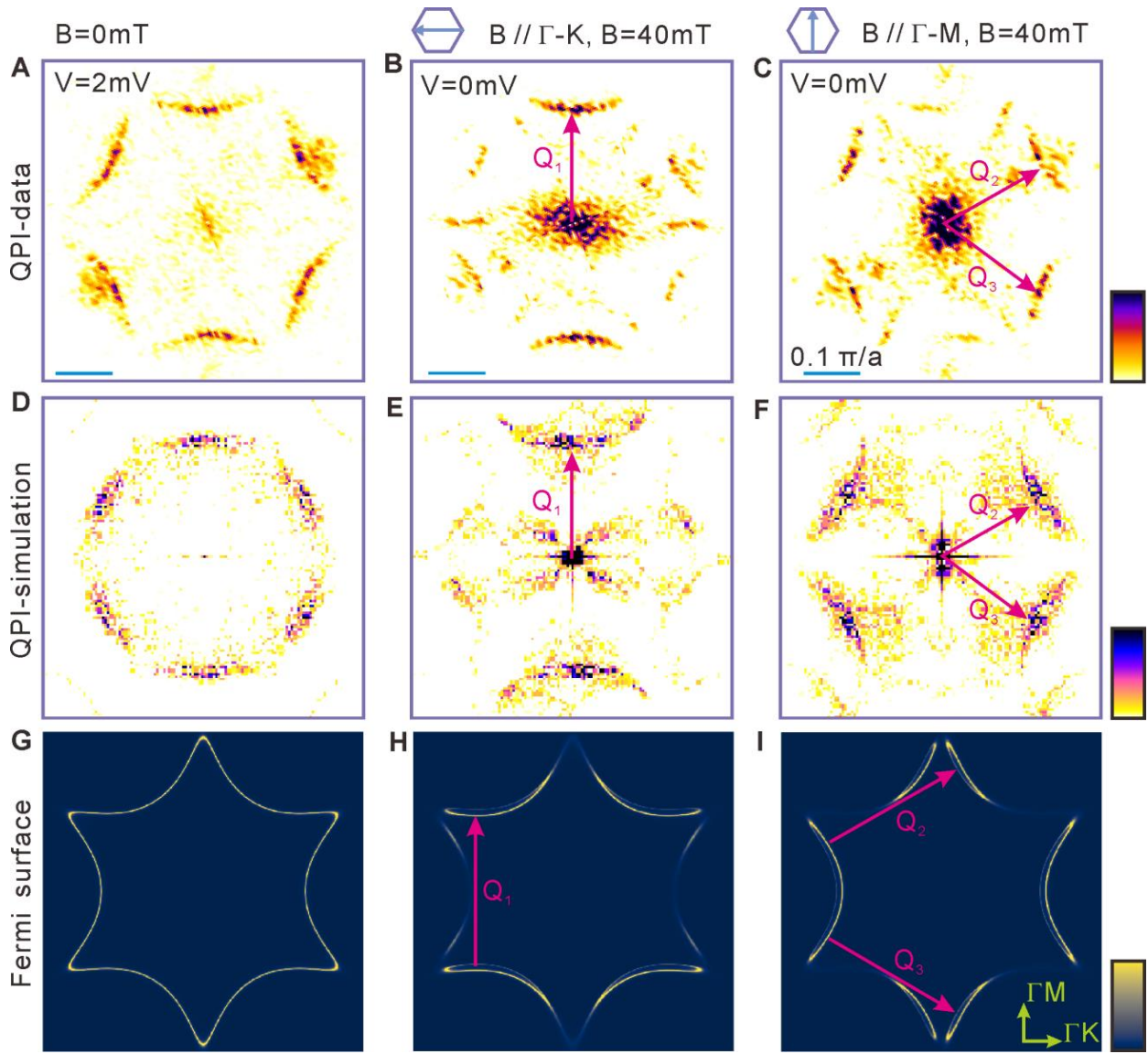

**Fig. 4 Quasiparticle interference patterns of segmented Fermi surface.** (A) QPI at zero magnetic field outside of superconducting gap. The six intensive segments correspond to scattering between the tip of the normal state Fermi surface. (B) QPI with B=40mT along ΓK at E=0. The two intensive segments correspond to vertical scattering between the tips of the star indicated in (H). (C) QPI with B=40mT along ΓM at E=0. The four intensive segments correspond to diagonal scattering between the tips of the star indicated in (I). (D-F) Numerical simulations for QPI in hexagonally warped disordered Dirac surface state corresponding to the magnetic fields as in (A-C). (G-I) Fermi surface contours corresponding to the magnetic fields as in (A-C)

# Supplementary Materials for

# Discovery of segmented Fermi surface induced by Cooper pair momentum

Zhen Zhu[1+], Michał Papaj[2+], Xiao-Ang Nie[1], Hao-Ke Xu[1], Yi-Sheng Gu[1], Xu Yang[1], Dandan Guan[1], Shiyong Wang[1], Yaoyi Li[1], Canhua Liu[1], Jianlin Luo[3], Zhu-An Xu[4], Hao Zheng[1*], Liang Fu[2*], Jin-Feng Jia[1*]

**Materials and Methods**

The study was performed using a commercial Ultra-high vacuum (UHV) system (Unisoku, USM1600), which consists of a molecular beam epitaxy (MBE) chamber and a scanning tunneling microscope (STM) equipped with a dilution refrigerator and a vector magnetic field ($H_x$:2T, $H_y$:2T, $H_z$:9T). $NbSe_2$ substrate was cleaved in UHV after being fully degassed. $Bi_2Te_3$ thin films were grown using elementary Bi and Te sources. The growth rate was calibrated by reflection high-energy electron diffraction (RHEED) intensity oscillations. After growth, the samples were transferred to STM and cooled under zero magnetic field down to 40mK. The effective electron temperature during the measurement is 300mK as determined by calibration using Pb, shown in Fig. S1. The in-plane magnetic field directions are determined by the in-plane vortex on $NbSe_2$, which is displayed in Fig. S2. The dI/dV signals were acquired by a lock-in amplifier with $V_{mod}$=50μV and f=991Hz. The dI/dV grid was obtained by measuring single dI/dV spectra at each spatial point with junction set points at 2mV, 1nA. The real space quasiparticle interference (QPI) pattern is the 0mV layer from the entire grid. To obtain the momentum space representation, we Fourier transform the real space data and show the absolute values of the results using linear scale.

**Supplementary Text**

<u>Experimental data for sample 2</u>

To further verify the results presented in the main text, we have prepared another $Bi_2Te_3$ thin film, also 4 quintuple layers thick, on top of $NbSe_2$, which we refer to as sample 2. Since our samples are very sensitive to low magnetic fields, we exhausted all liquid helium and warmed up our entire cryostat to about 30K in order to minimize the residual magnetic fields after the measurements of sample 1. The dI/dV spectra of sample 2 (Fig. S3) we observe are qualitatively similar to those of sample 1, especially with regard to the presence of multiple in-gap features that vary with the magnetic field. For example, a peak at zero bias appears at 50mT when the field is applied along Γ-M direction, which is slightly higher compared to sample 1, where it appears at 40mT as displayed in Fig. 2. The discrepancy may be associated with the accuracy of our magnetic field controller. The magnetic coils in our cryostat can generate a field of up to 2T in x and y directions, which requires a large current of over 100A. At the low field limit, we can only control the field magnitude down to every 10mT. Therefore, the field intensities that generate a peak at zero energy are of only one step difference between both samples. The reproducibility of the observed in-gap features validates our MBE growth recipe and allows us to repeatably prepare uniform $Bi_2Te_3/NbSe_2$ thin films of high quality.

<u>QPI measurements with and without magnetic field</u>

To quantitatively determine the length of the scattering vector, we measure the QPI pattern at the energy outside of the superconducting gap under zero magnetic field with both the scattering features and the Bragg points visible. In Fig. S4, we show an intensity line profile along the QPI map and find that the scattering vector $Q \approx 0.1 \times 2\pi/a$, where $a$ is the lattice constant of $Bi_2Te_3$(111) surface. On the surface of our 4QL $Bi_2Te_3$/$NbSe_2$ heterostructure under zero field, the dI/dV map measured at the energy outside of superconducting gap demonstrates clear standing wave pattern. In contrast, the dI/dV map taken at zero energy (center of the gap) is featureless. Fig. S5 shows the real space and reciprocal space QPI patterns.

Based on the results in main text, we know that in-plane magnetic field produces distinct in-gap features and that the in-gap QPI pattern depends strongly on the orientation of the magnetic field. In contrast, in Fig. S6, one can clearly observe that the QPI for energies outside of the gap under a field of 40mT along either Γ-K or Γ-M direction displays no obvious difference as compared to zero field QPI at the same energy. It further demonstrates that the anisotropic QPI patterns are the result of the in-gap quasiparticle state scattering. The data represented in Figs. S4-S6 are all acquired on the sample 1, which is the same sample as discussed in the main text.

Density of states and QPI on the pristine surface of $NbSe_2$

To further demonstrate that the standing wave patterns observed in the sample presented in the main text indeed arise due to the $Bi_2Te_3$ thin film, we perform additional experiments on a pristine surface of $NbSe_2$. As shown in Fig. S7, we measure dI/dV spectra on $NbSe_2$ surface under an in-plane magnetic field with increasing intensities up to 80mT. No significant in-gap features can be observed apart from a small change to the coherence peaks. Moreover, the dI/dV map at 0V and its corresponding FFT show no QPI signals. Moreover, for field magnitudes discussed in the main text we have not detected any vortices in the measurement field of view. These observations taken together further prove that the features discussed in main text do not come from the $NbSe_2$ substrate.

Continuum model for density of states calculation

The density of states calculation is performed using a continuum model defined by a Bogoliubov-de Gennes Hamiltonian. The model describes a topological Dirac surface state with hexagonal warping and s-wave superconducting order parameter introduced via proximity effect:

$$H(\mathbf{k}) = v(k_x\sigma_y - k_y\sigma_x)\tau_z + \lambda(k_x^3 - 3k_xk_y^2)\sigma_z\tau_z - \mu\tau_z + \Delta\tau_x,$$

where $\sigma_i, \tau_i$ are Pauli matrices describing the spin and particle-hole degrees of freedom, respectively, $v$ is the velocity of the Dirac surface state, $\lambda$ is the parameter determining the strength of hexagonal warping, $\mu$ is the chemical potential and $\Delta$ is the superconducting order parameter. The magnetic field is included in this Hamiltonian via the minimal coupling substitution $k_i \to k_i - eA_i\tau_z$ with $A_i$ being the components of the vector potential. We then calculate the spectral function of this model:

$$A(\mathbf{k}, \omega) = -\frac{1}{\pi} Im\left(G_{11}^R(\mathbf{k}, \omega) + G_{22}^R(\mathbf{k}, \omega)\right), \qquad G^R(\mathbf{k}, \omega) = \left(\omega + i\eta - H(\mathbf{k})\right)^{-1}$$

and obtain the density of states via numerical integration $\rho(\omega) = \int_{\mathbf{k}} A(\mathbf{k}, \omega)$, normalized by the density of states outside of the superconducting gap. The amount of hexagonal warping can be characterized by the ratio of the characteristic energy scale of the warped normal state, $E^* = v\sqrt{v/\lambda}$ and the chemical potential. In the calculation for Fig.2 we used $E^*/\mu = 1.6$ and $\mu/\Delta = 160$. The magnetic field is applied in two perpendicular directions that correspond to ΓM and ΓK axes of the Brillouin zone.

To demonstrate that the presence of distinct in-gap features observed in the experiment is enabled by the hexagonal warping of the topological surface state, we contrast the density of states results presented in the main text with the calculation for $\lambda = 0$. First of all, in such a case the spectra are rotationally invariant with respect to the orientation of the external magnetic field. Second, as shown in Fig. S8, the evolution of density of states is qualitatively different, with neither in-gap peaks nor multiple shoulders appearing.

The origin of the in-gap features can be understood by looking more closely at the spectral functions corresponding to different energy values. To visualize this, we present two sample animations for magnetic field along ΓM and ΓK directions, which show the evolution of the spectral function as a function of energy. The peaks and shoulders correspond then to the appearance and disappearance of quasiparticle Fermi surface pockets along the hexagonally warped contour.

<u>Vector potential near the surface of a superconductor</u>

To estimate the magnitude of the vector potential in the vicinity of a superconductor due to the in-plane magnetic field, we follow a standard calculation based on the London equation:

$$\boldsymbol{j}_S = -\frac{n_S e^2}{m^*} \boldsymbol{A}$$

where $\boldsymbol{j}_S$ is the supercurrent density, $n_S$ is the superconducting electron density and $m^*$ is the effective electron mass. As we are working in the London gauge ($\nabla \cdot \boldsymbol{A} = 0$), we can use the Maxwell equations to get:

$$\nabla^2 \boldsymbol{A} = \frac{1}{\lambda^2} \boldsymbol{A}$$

with $\lambda = \sqrt{\frac{m^*}{\mu_0 n_S e^2}}$ being the London penetration depth. Choosing the magnetic field to be along y direction, we can look for the solution of the form:

$$\boldsymbol{A} = A_x(z)\hat{x}$$

Therefore, we are now solving:

$$\partial_z^2 A_x = \frac{1}{\lambda^2} A_x$$

in the geometry of a rectangular superconducting slab placed within $-h_{SC} < z < 0$ and with the boundary condition giving the external magnetic field $B_0$ outside of the superconductor:

$$B_y(z = -h_{SC}) = \partial_z A_x(z = -h_{SC}) = B_0, \qquad B_y(z = 0) = \partial_z A_x(z = 0) = B_0$$

In such a case the solution within the superconductor is:

$$A_x(z) = \frac{B_0 \lambda \sinh\left(\frac{z}{\lambda} + \frac{h_{SC}}{2\lambda}\right)}{\cosh\frac{h_{SC}}{2\lambda}}, \qquad -h_{SC} < z < 0$$

Outside of the superconductor, inside of the thin film of the topological insulator, vector potential can be obtained from continuity:
$$A_x(z) = B_0 z + B_0 \lambda \tanh\left(\frac{h_{SC}}{2\lambda}\right), \quad z > 0$$
As the thickness of the thin film $h_{TI}$ is much smaller than $\lambda$ and $h_{SC} \gg \lambda$, we can approximate the vector potential at the top surface of the film to be:
$$A_x(z = h_{TI}) \approx B_0 \lambda$$

At 20 mT, the approximate field of gap closing, this corresponds to $\delta k = \frac{eB_0\lambda}{\hbar} \approx 3 \cdot 10^{-4}$ Å$^{-1}$ with $\lambda \approx 100$ nm. Comparing this to Fermi wavevector measured using ARPES $k_F = 0.17$Å$^{-1}$ we have $\frac{\delta k}{k_F} \approx 1.7 \cdot 10^{-3}$. As a sanity check, we can compare this ratio to the ratio of superconducting gap and the chemical potential. From the dI/dV curves we have $\Delta \approx 0.5$ meV and $\mu = 360$ meV, so $\frac{\Delta}{\mu} \approx 1.4 \cdot 10^{-3}$, so the gapless superconducting state arises when the energy scale associated with the momentum shift is of the order of the superconducting gap as expected. Such phenomenon was also proposed as a source of enhanced g-factor in quantum spin Hall insulators(29).

Tight-binding model for quasiparticle interference pattern simulations

For the QPI simulations we use a real space approach based on recursive Green's functions algorithm as discussed in the context of high temperature superconductors(30). In this method the calculation is performed in a cylindrical geometry with hard-wall boundary conditions in the x direction and periodic boundary conditions in the y direction. To minimize the effect of the boundaries, we add an exponentially decaying imaginary part to the diagonal of the Hamiltonian on each orbital and lattice site of the form:
$$i\,\eta_B(x,y) = \eta_0 \left(\exp\left(-\frac{x}{l_x}\right) + \exp\left(-\frac{L-x}{l_x}\right) + \exp\left(-\frac{y}{l_y}\right) + \exp\left(-\frac{W-y}{l_y}\right)\right)$$
In the simulations we choose a model with a single hexagonally warped Dirac cone in the Brillouin zone. To realize this scenario, we use a honeycomb lattice with nearest and next-nearest neighbor hoppings (Fig. S9). In this representation the two basis elements correspond to the two spin components of the Dirac fermion. The nearest neighbor hoppings $it_1$ result in graphene-like Dirac cones in K and K' points of the Brillouin zone. Addition of the next nearest neighbor complex hoppings $t_2 e^{\pm i\phi}$ with alternating phase sign breaks time-reversal symmetry and in this way gaps out one of the two Dirac cones while introducing the hexagonal warping to the remaining one when $\phi=7/6\pi$. The superconductivity is then introduced by doubling the number of degrees of freedom at each lattice site and coupling them by the superconducting gap order parameter $\Delta$. Finally, the magnetic field appears as the spatially dependent phase of the superconducting order parameter $\Delta(\mathbf{r}) = \Delta \exp\left(i\frac{2e}{\hbar}\mathbf{A}\cdot\mathbf{r}\right)$. With such a tight-binding model in place we calculate the local density of states for the whole system, taking into account that both of the spin components reside at the same point in real space. To introduce scattering we add random on-site potential at each of the lattice sites, chosen from the range $[-V_0/2, V_0/2]$. The parameters used in the simulations are $t_1 = 2/\sqrt{3}$, $t_2 = 16$, $\mu = 0.8$, $\Delta = 0.04$, $V_0 = 0.1$, $\eta_0 = 0.5$, $l_x = 30$, $l_y = 8$.

The lattice size was chosen to be $L = 620$, $W = 460\frac{\sqrt{3}}{2}$ with lattice constant $a = 1$. The QPI pattern is then obtained by performing the Fourier transform on the real space results taken from a central square with side length equal to 298.

Additional details for the structure of the QPI pattern under magnetic field

To fully elucidate the origin of the anisotropic in-gap QPI patterns under magnetic field, it is necessary to modulate the joint density of states at the initial and final scattering states by the superconducting coherence factors that carry the information about the spin-momentum locking of the topological surface state, and the mixing of particle/hole components. In order to better visualize this, we calculate the electron spectral function in momentum space as described in the previous section and then filter it by the matrix elements of the point impurity scattering potential between a fixed point $\boldsymbol{k}_0 = (k_{x0}, k_{y0})$ and the remaining states:

$$\Gamma(\boldsymbol{k}) = A(\boldsymbol{k}, \omega_0) \left( |\langle \psi_1(\boldsymbol{k})|V_{imp}|\psi_0(\boldsymbol{k}_0)\rangle|^2 + |\langle \psi_2(\boldsymbol{k})|V_{imp}|\psi_0(\boldsymbol{k}_0)\rangle|^2 \right)$$

where $V_{imp}$ is independent of momentum, $\psi_0(\boldsymbol{k}_0)$ is the chosen initial state wave function at the energy $\omega_0$, and $\psi_i(\boldsymbol{k})$ are the two eigenstates of the Bogoliubov-de Gennes Hamiltonian relevant at $\omega_0$ when $\mu$ is the largest energy scale of the problem and we can only focus on the upper Dirac cone of the topological surface state. In Fig. S10 we present a set of spectral functions originally shown as insets of Fig. 4 of the main text, together with the transition rates $\Gamma(\boldsymbol{k})$ determined by the wave function overlaps with the state at $\boldsymbol{k}_0$, marked by the points on the spectral function panels. We observe then that while the normal state pattern of six equal segments arises from scattering between the tips of the Fermi surface star, when superconductivity is included some of these scattering processes are strongly suppressed, resulting in diminished segments in the QPI pattern.

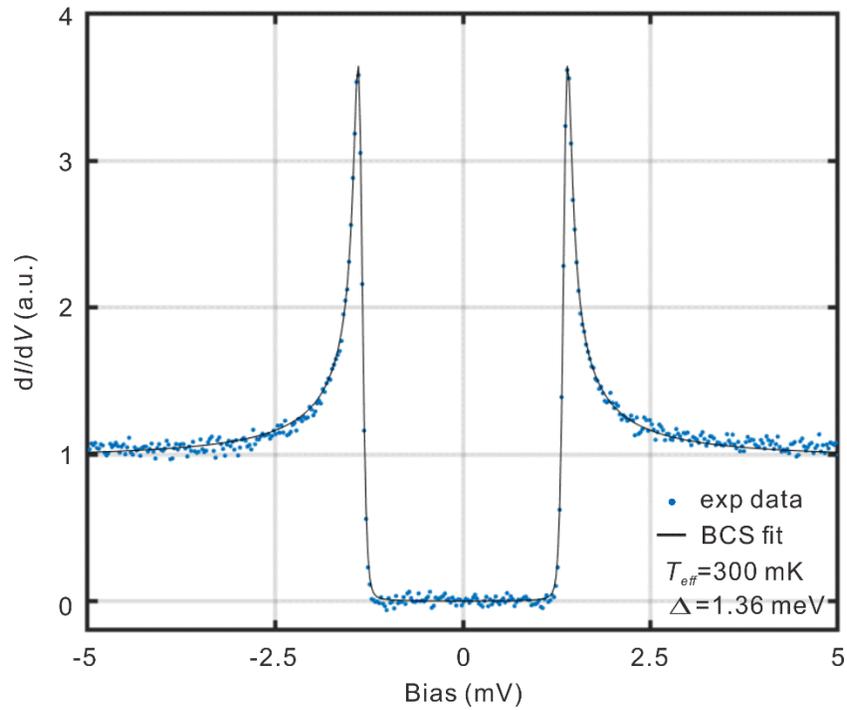

**Fig. S1.**
**Superconducting gap and BCS fitting on Pb.** The dI/dV spectrum was acquired on a polycrystalline Pb foil with a clean W tip at a sample temperature of T = 40 mK. The set point is 5mV, 1nA. Blue dots are the experimental data. Black curve is the BCS fit with Δ=1.36 meV, which yields an effective electron temperature $T_{eff}$ = 300 mK.

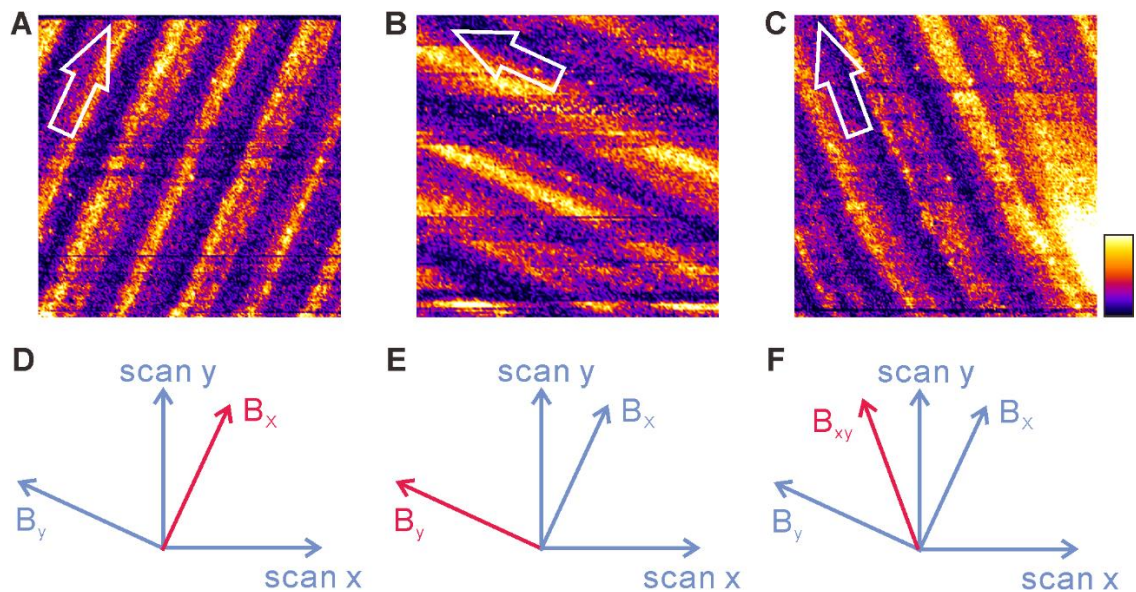

**Fig. S2.**
**Calibration of the field directions.** (**A**)-(**C**) dI/dV maps that display the in-plane vortices on NbSe$_2$ surface. Image size: 300nm × 300nm. The maps are the 0mV layers of the dI/dV grids, for which set points are 3mV and 0.1nA. The magnetic field of 0.5T is applied to the samples with directions indicated by the white arrows. In-plane field induced vortices manifest as bright stripes in the map. (**D**)-(**F**), relations between the field directions and the scan frame.

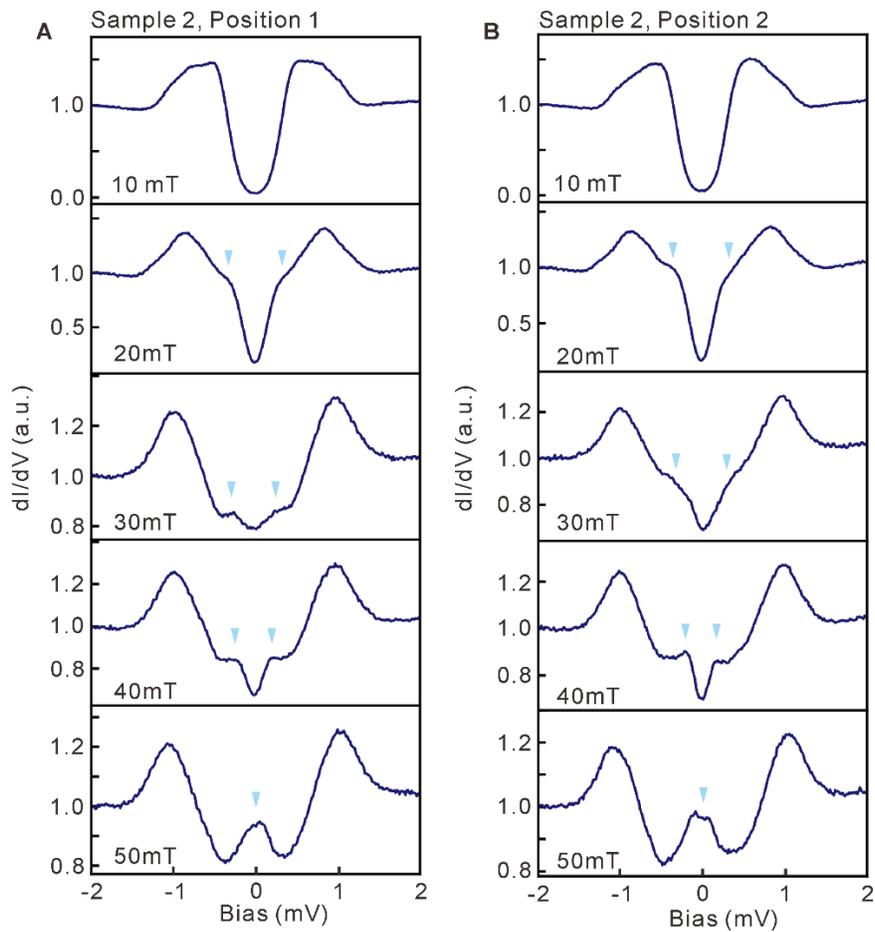

**Fig. S3.**
**Field induced in-gap states in sample 2.** (**A**) and (**B**) are dI/dV spectra measured at two positions of sample 2 under various in-plane magnetic fields with direction parallel to Γ-M. Set point: V=2mV, I=0.5nA, T=40mK. Blue arrows point to the characteristic in-gap quasiparticle features.

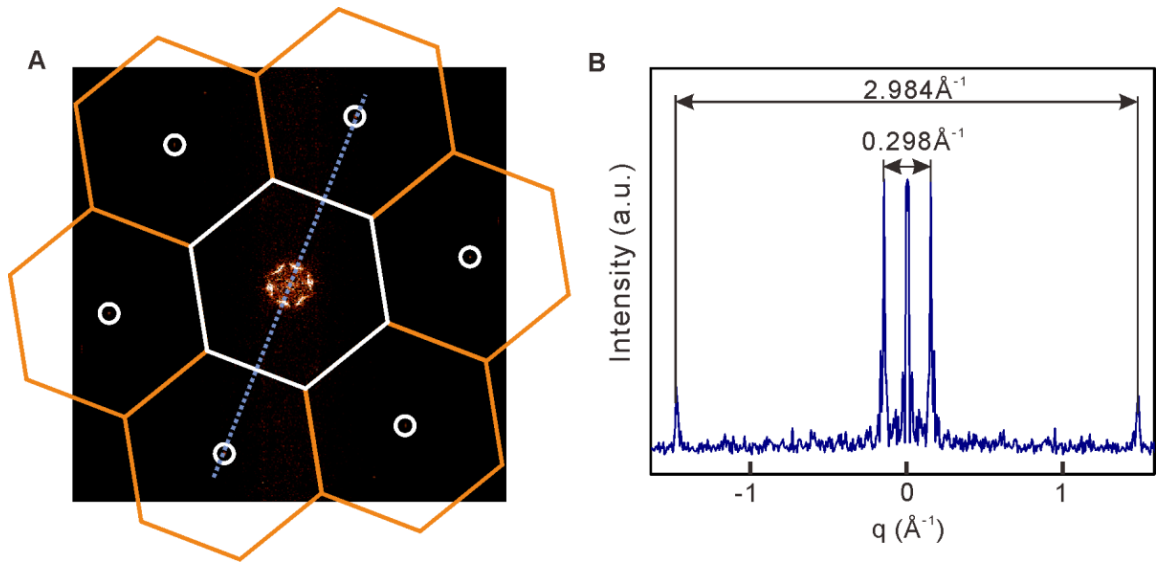

**Fig. S4.**
**Large-scale QPI pattern together with Bragg points.** (**A**), QPI measured under zero field in a large-scale reciprocal space. Six Bragg points are marked with white circles. The 1st and 2nd Brillouin zones are labelled with white and orange hexagons, respectively. Set point is 3mV, 0.5nA. Note that the arcs of the QPI features are parallel to the zone boundaries. (**B**), Intensity profile along the blue dashed line in A. Based on this measurement, we determine scattering vector, $Q \approx 0.1 \times 2\pi/a$, where $a$ is the in-plane lattice constant of $Bi_2Te_3$.

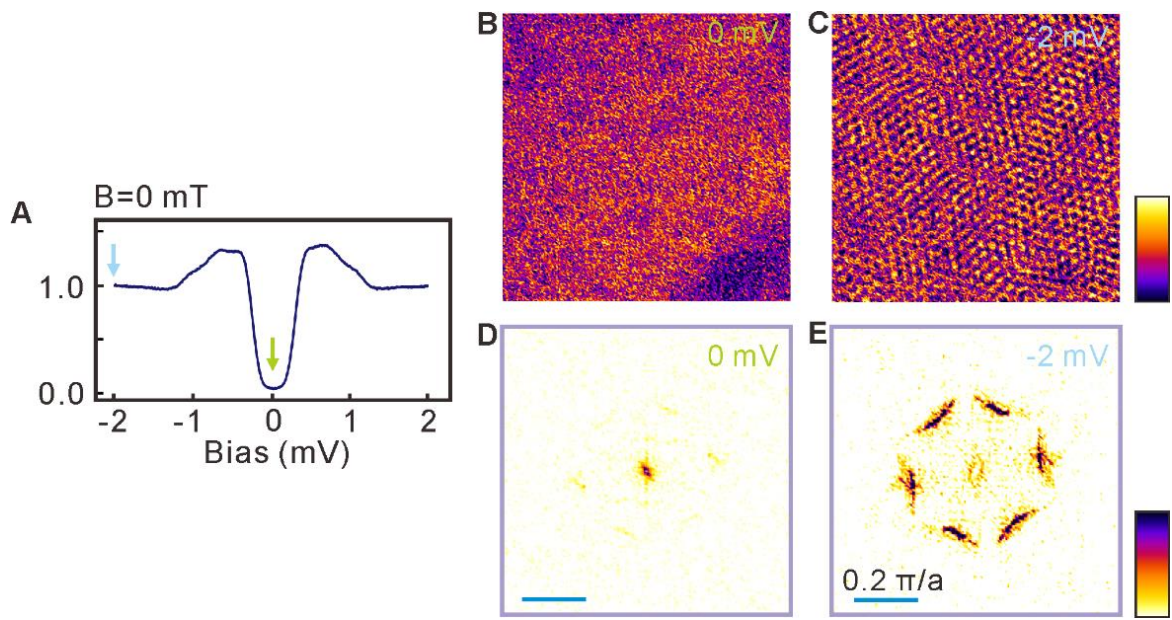

**Fig. S5.**
**Spectroscopy and QPIs at zero field.** (**A**), dI/dV spectra measured under zero field. Hard superconducting gap induced by proximity effect is present. Set point is 2mV, 0.5nA. (**B**) and (**C**), dI/dV maps taken at 0mV (green arrow in A) and -2mV (blue arrow in A) under zero field. Noteworthy standing wave patterns are resolved at -2mV. (**D**) and (**E**), the Fourier transforms of B and C. QPI measured at 0mV shows negligible signal in contrast to the -2mV one with six equivalent arcs.

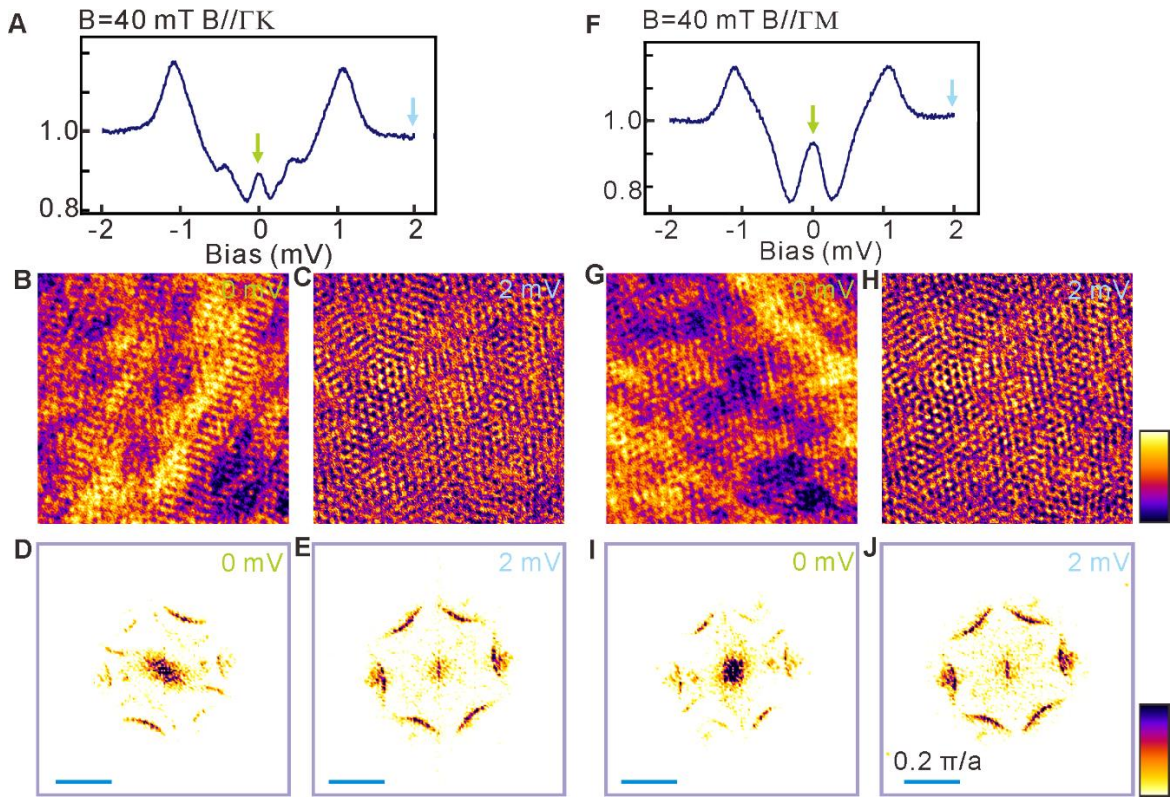

**Fig. S6.**

**QPIs of in-gap and off-gap states under in plane field.** (**A**), dI/dV spectra measured under 40mT with field direction parallel to Γ-K. (**B**) and (**C**), dI/dV maps taken at 0mV (green arrow in A) and 2mV (blue arrow in A) with the same magnetic field as in A. (**D**) and (**E**), the Fourier transforms of B and C. QPI measured at 2mV shows six equivalent arcs, compared to the 0mV one with two dominating arcs. (**F**), dI/dV spectra measured under 40mT and along Γ-M. (**G**) and (**H**), dI/dV maps taken at 0mV (green arrow in F) and 2mV (blue arrow in F) with the same magnetic field as in F. (**I**) and (**J**), the Fourier transforms of G and H. QPI measured at 2mV shows six equivalent arcs, as compared to the 0mV one with four dominating arcs. We note there is no noticeable difference for the QPI pattern measured at 2 mV (outside of the gap) between the two magnetic field orientations in E and F.

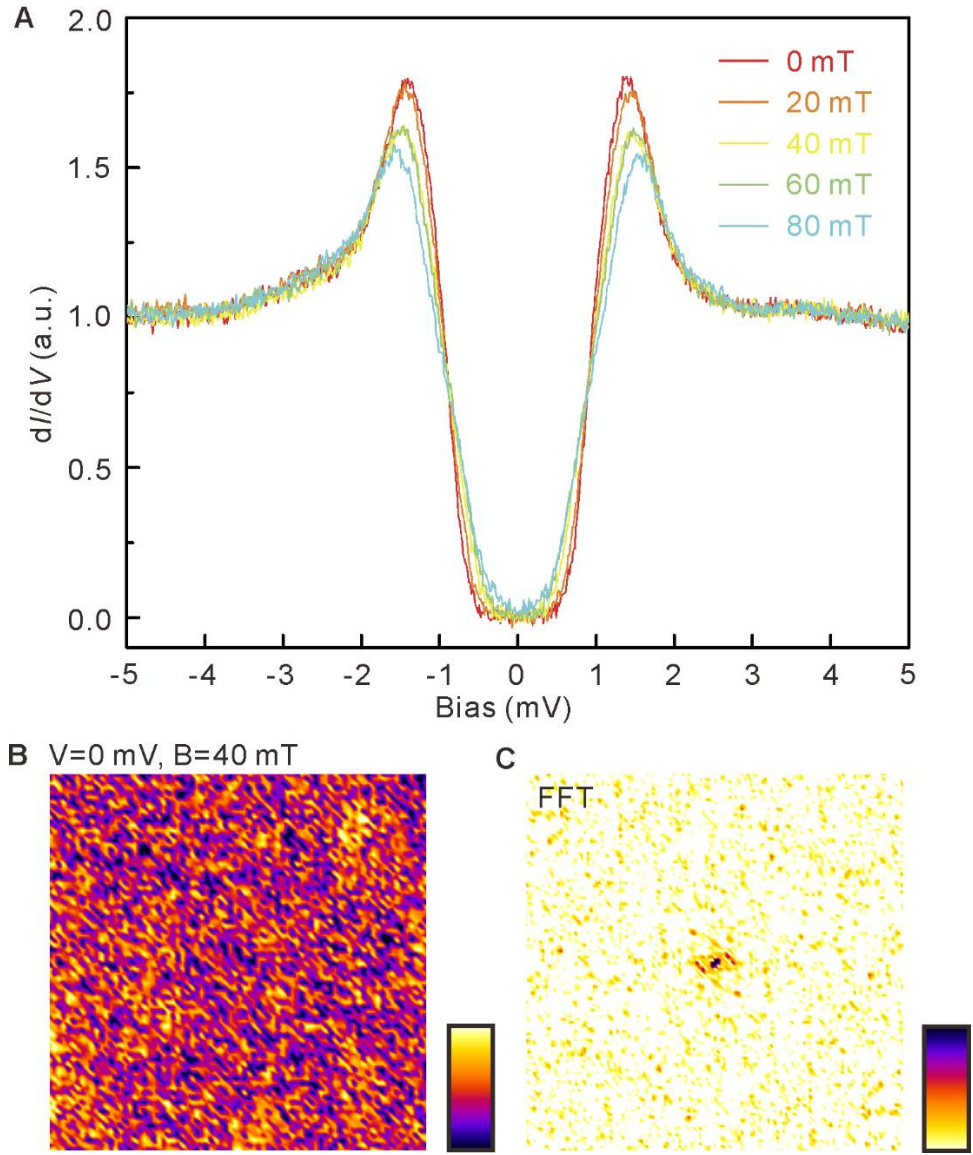

**Fig. S7.**
**dI/dV spectra and QPI on the pristine surface of NbSe₂ under in-plane magnetic field**. (A) dI/dV spectra measured on NbSe$_2$ under increasing magnetic field. No in-gap features are observed, and the superconducting gap remains hard. The set point is 5mV, 0.1nA. (**B**) and (**C**) are dI/dV maps (120nm×120nm, 3mV, 0.1nA) taken at 0mV and the Fourier transform of the dI/dV map, respectively. Instead of standing waves observed for Bi$_2$Te$_3$ at 40mT, the signal measured on NbSe$_2$ mainly originates from noise, which is consistent with the zero DOS at the Fermi level in A. We also do not detect any vortices in the measurement field of view under such magnetic field

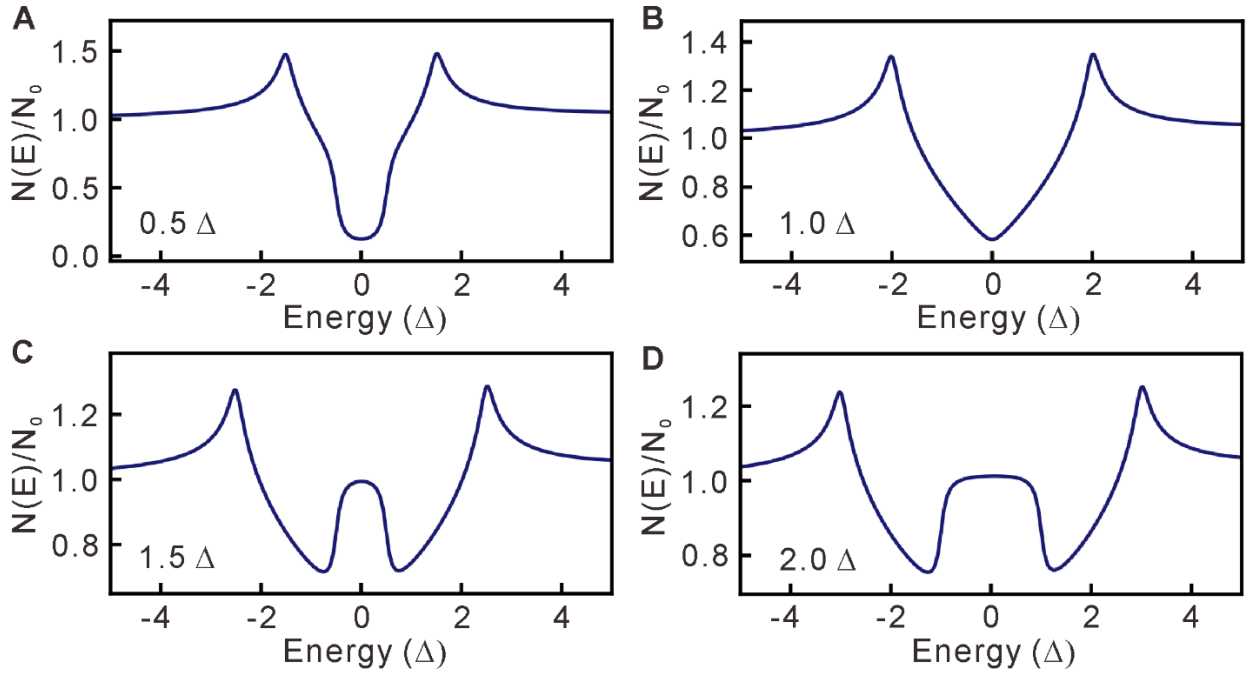

**Fig. S8.**
**Calculations of DOS without hexagonal warping effect.** (**A**)-(**D**) Plots of theoretical curves of DOS for a proximitized ideal 2D helical Dirac gas under various in-plane magnetic fields. The field intensities are represented by the Zeeman energies in the units of superconducting gap $\Delta$. The isotropic nature of the Dirac cone makes the resulting density of states independent of magnetic field direction.

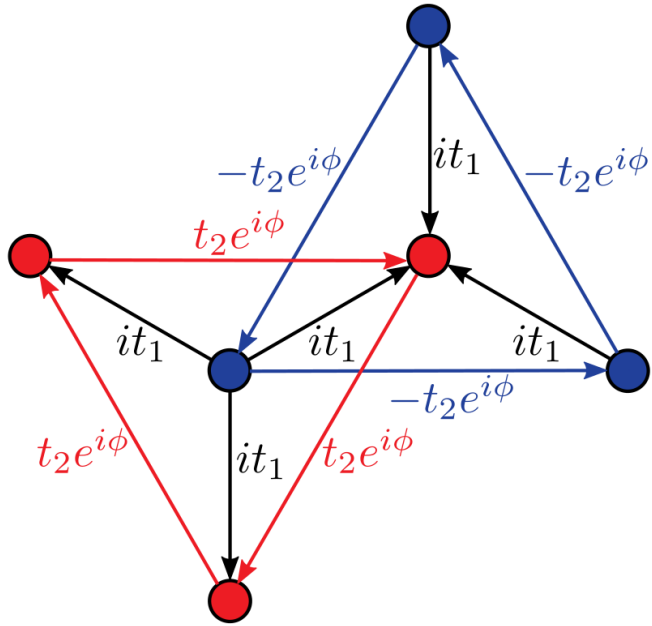

**Fig. S9.**
**The tight-binding model for QPI simulations.** Visualizations of the honeycomb lattice model with all the relevant hoppings. Sublattices correspond to the two spin components. Each lattice site has particle and hole degrees of freedom that are coupled to introduce superconductivity.

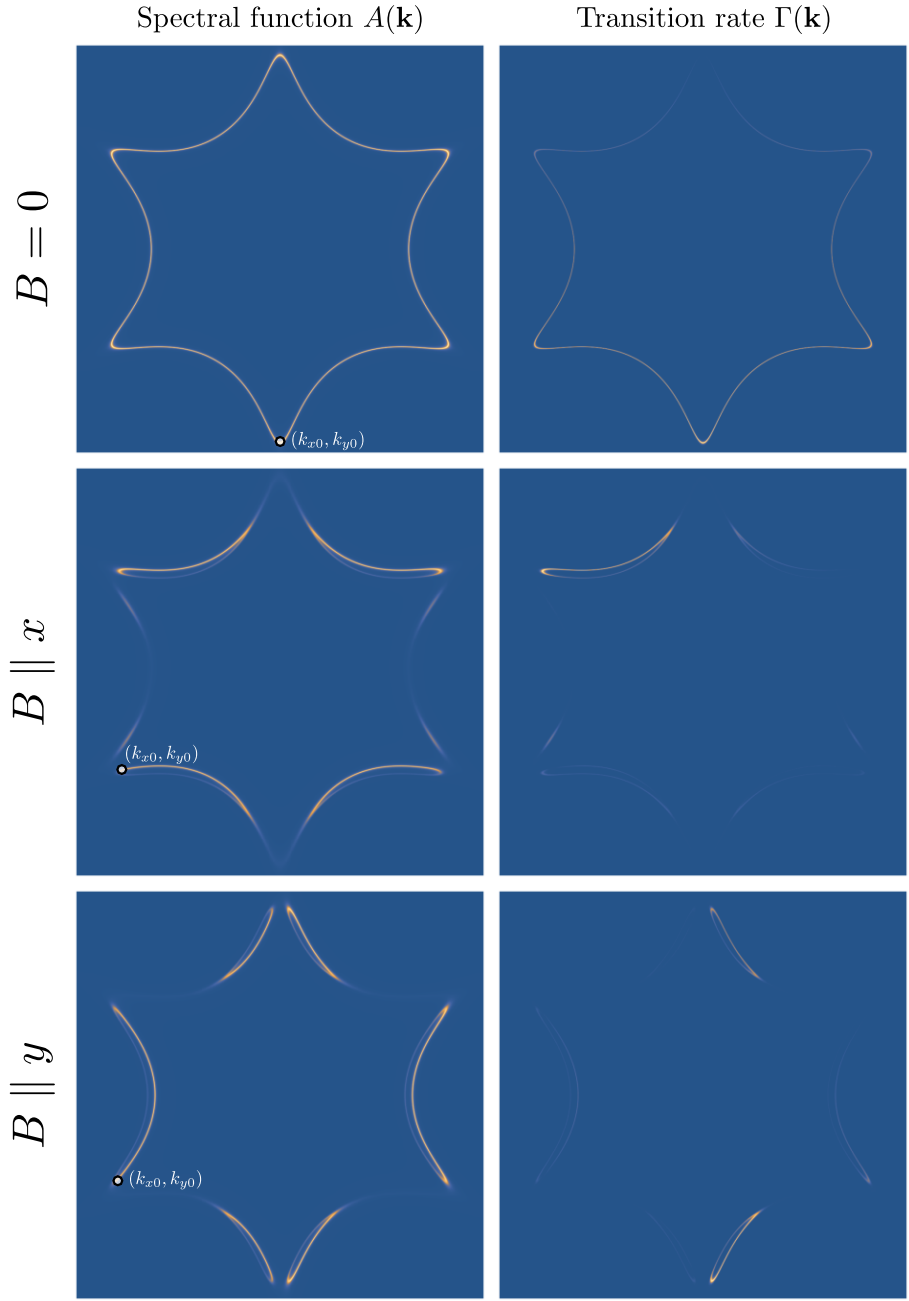

**Fig. S10.**

**The transition rates for scattering from the Fermi surface star tips.** Spectral functions $A(\boldsymbol{k})$ (left column) of the quasiparticle Fermi surfaces at $E = 0$ with and without magnetic field. $B = 0$ row displays the spectral function of the normal state. Right column shows the transition rate $\Gamma(\boldsymbol{k})$ with matrix elements calculated with respect to the initial states marked by the white points in the spectral function panels. This displays the suppression of scattering to some of the neighboring tips of the Fermi surface in superconducting state under in-plane magnetic field. This results in disappearance of some of the segments visible in the QPI pattern of the normal state.